# Elastic Anomalies Associated with the Antiferroelectric Phase Transitions of PbHfO$_3$ Single Crystals

Jae-Hyeon Ko*

*Department of Physics, Hallym University, Chuncheon, Gangwondo 200-702*

Krystian Roleder

*Institute of Physics, University of Silesia, ulica Uniwersytecka 4, PL-40-007 Katowice, Poland*

Andrzej Majchrowski

*Institute of Applied Physics, Military University of Technology, ul. Kaliskiego 2, 00-908 Warsaw, Poland*

Annette Bussmann-Holder

*Max-Planck-Institut für Festkörperforschung, Heisenbergstr. 1, D-70569 Stuttgart, Germany*

The temperature dependence of the elastic properties of antiferroelectric PbHfO$_3$ was investigated by Brillouin scattering. The two structural phase transitions of antiferroelectric-antiferroelectric-paraelectric phases were clearly identified by discontinuous changes in the acoustic mode frequencies and the hypersonic damping. The substantial softening of the mode frequency along with the remarkable increase in the acoustic damping observed in the paraelectric phase indicated the formation of precursor noncentrosymmetric (polar) clusters and their coupling to the acoustic waves. This was corroborated by the observation of quasi-elastic central peaks, the intensity of which grew upon cooling toward the Curie point. The obtained relaxation time exhibited a slowing-down behavior, suggesting that the dynamics of precursor clusters becomes more sluggish on approaching the phase transition temperature.

PACS numbers: 78.35.+c, 62.50.+p

Keywords: Lead hafnate, $PbHfO_3$, Brillouin scattering, elastic properties, antiferroelectrics

*Email: hwangko@hallym.ac.kr

Fax: +82-33-256-3421

# I. INTRODUCTION

Antiferroelectricity has attracted great attention recently not only due to its potential applicability in various fields but also due to its relation to the fundamental physics of phase transitions [1]. Antiferroelectrics are promising materials for applications as actuators, energy-storage capacitors, and some devices using giant electrocaloric effects [2]. $PbZrO_3$ is the most representative antiferroelectric perovskite oxide. Recent lattice dynamic studies on this compound revealed rich information associated with the nature of the antiferroelectric phase transition [3-6]. In spite of the fact that the microscopic nature of the antiferroelectric phase transition in $PbZrO_3$ is still under debate, it seems evident that more detailed and thorough investigations into the lattice dynamics over the whole momentum space in the Brillouin zone are indispensable for our better understanding of this archetypal antiferroelectrics.

Another interesting antiferroelectric materials is lead hafnate, $PbHfO_3$, which has been less studied as compared to $PbZrO_3$. $PbHfO_3$ exhibits two successive antiferroelectric phase transitions at ~ 210 ºC and ~ 160 ºC [7-8]. Large thermal vibrations of Pb cations and some signature of dynamic or static disorder were revealed in the paraelectric phase [9-10]. Detailed x-ray and neutron scattering studies revealed that the low-temperature phase down to 10 K is centrosymmetric *Pbam* isomorphous to the antiferroelectric phase of $PbZrO_3$ [11-12]. Recent synchrotron and neutron powder diffraction studies showed that Pb cations are disordered in the cubic phase of $PbZrO_3$ and $PbHfO_3$ and that thermal vibrations of oxygen atoms exhibit strong anisotropy [13-15]. All these studies indicate that the motions of cations and oxygen ions in $PbHfO_3$ are very anharmonic in the cubic phase, which in some way must be related to the precursor phenomena of the antiferroelectric phase transition.

In spite of previous studies on $PbHfO_3$, fundamental material properties such as elastic characteristics of this compound have not been reported yet due to the lack of high-quality single crystals. We report for the first time the elastic properties of $PbHfO_3$ studied by Brillouin scattering. Previous Brillouin scattering study on $PbZrO_3$ revealed substantial acoustic mode softening and associated precursor phenomena in the centrosymmetric cubic phase [4]. This study is focused on the precursor phenomena of PbHfO$_3$ and the associated elastic anomalies, which will be compared to that of PbZrO$_3$.

## II. EXPERIMENTAL SETUP

PbHfO$_3$ single crystals were grown by means of spontaneous crystallization from high temperature solution in Pb$_3$O$_4$-B$_2$O$_3$ solvent. The composition of the melt used in our experiments was the same as in Ref. 16 devoted to crystallization of PbZrO$_3$, namely 2.4 mol% of PbHfO$_3$, 77 mol% of PbO (re-counted to Pb$_3$O$_4$) and 20.6 mol% of B$_2$O$_3$. Pb$_3$O$_4$ was used instead of PbO to avoid coloration of the as-grown crystals caused by oxygen deficiency. The crystallization was carried out in a platinum crucible covered with a platinum lid under conditions of low temperature gradient. After soaking at 1473K for 24 hours the melt was cooled to 1200K at a rate of 3.5 K/h and after decantation the furnace was cooled to the room temperature at a rate of 10 K/h. As-grown PbHfO$_3$ single crystals were etched in diluted acetic acid to remove residues of the solidified flux.

Brillouin spectrum was measured by using a conventional tandem six-pass Fabry-Perot interferometer (TFP-1, JRS Co.). A backscattering geometry was adopted by using a microscope (Olympus BX-41). A compact cryostat (THMS600, Linkam) was put on the microscope stage for temperature control. A diode-pumped solid state laser (Excelsior 532-300, SpectraPhysics) at a wavelength of 532 nm was used as an excitation source. Two free spectral ranges were used to cover a wide frequency range for probing the longitudinal acoustic (LA) mode propagating along the [100] direction and a quasi-elastic central peak. The details of the experimental setup can be found elsewhere [17-19].

## III. RESULTS AND DISCUSSION

Figure 1 shows a few selected Brillouin spectra in a frequency range of ±50 GHz. One Brillouin doublet, corresponding to the LA mode propagating along the [100] direction, is seen at 200 °C in the cubic phase. This is consistent with the Brillouin selection rule [20]. The LA mode frequency shifts to the lower frequency range and its line width increases upon cooling toward the transition temperature. When the PbHfO$_3$ undergoes a paraelectric-antiferroelectric phase transition at $T_H$~210 °C, the LA mode shows a splitting and the transverse acoustic (TA) mode appears suddenly. The LA mode width also shows a substantial change during the phase transition. LA and TA modes were fitted by using a Voigt function, where the Gaussian line width was fixed to that of the instrumental function of the interferometer.

Figure 2 shows the temperature dependence of the LA mode frequency ($\nu_B$) and the full width at half maximum (FWHM, $\Gamma_B$). $\nu_B$ exhibits a softening from ~45.5 GHz to ~39.5 GHz upon cooling in the cubic phase. This mode softening is accompanied by a large increase in $\Gamma_B$. In particular, the growth of $\Gamma_B$ becomes more substantial at temperatures below approximately 300 °C. The paraelectric-antiferroelectric phase transition at $T_H$ is characterized by a minimum of $\nu_B$ and a sharp maximum of $\Gamma_B$. The anomalous changes in both $\nu_B$ and $\Gamma_B$ reflect coupling between the LA waves and other dynamic degrees of freedom, the characteristic frequency of which is in or close to the hypersonic region. Similar mode softening of the LA waves has also been observed from other perovskite ferroelectrics and antiferroelectrics, such as BaTiO$_3$ and PbZrO$_3$ [4,21-22]. The LA mode exhibits splitting over the whole (intermediate) antiferroelectric phase. This may be due to the formation of multi domain structure and/or birefringence effect. The LA mode frequency shows a discontinuity at the antiferroelectric-antferroelectric phase transition at $T_L$~165 °C. This low-temperature phase transition is not accompanied by any abrupt change in $\Gamma_B$. The TA mode frequency, which is shown in the inset of Fig.2 (a), exhibits a discontinuous change at $T_L$ and this mode disappears exactly at $T_H$ and cannot be seen in the paraelectric phase. Its line width increases upon heating toward $T_H$.

Figure 3 compares the LA mode behaviors of $PbHfO_3$ and $PbZrO_3$. Both antiferroelectric perovskites display significant mode softening in the cubic phase along with the increase in the line width. However, there are two noticeable differences between these two materials. First, the mode frequency of $PbHfO_3$ is systematically lower than that of $PbZrO_3$. There has been no report on the first-principle calculations on $PbHfO_3$ to the authors' knowledge. According to the charge density distribution determined by Rietveld analysis[13], the Pb cations are rather ionic rather than showing covalency. The differences in the bonding nature and charge distributions may cause the difference in the elastic constants between these two antiferroelectrics. Also, $\nu_B$ shows a discontinuous jump in $PbZrO_3$ while this discontinuity in $\nu_B$ is small in the case of $PbHfO_3$. This might indicate that disorder in Pb sublattice in $PbHfO_3$ (i.e. an amplitude of Pb shift from the ideal crystallographic positions defined by Pm3m symmetry for these ions) is not so strong as that in $PbZrO_3$.

In $BaTiO_3$ and $PbZrO_3$, the mode softening above the phase transition was attributed to the formation of precursor polar regions and their coupling to the acoustic modes [4,21-22]. Lattice instability and mode coupling have been considered to play an important role in the phase transition of polar dielectrics [23]. The dynamic behavior of precursor polar clusters manifests themselves as quasi-elastic central peaks centered at zero frequency [4,24-25]. Figure 4 shows the temperature variation of the quasi-elastic central peak measured in the frequency range of ±560 GHz, corresponding to the wavenumber of approximately ±18 cm$^{-1}$. Compared to the spectrum at the highest measurement temperature of 600 °C, the central peak grows significantly upon cooling toward $T_H$. In addition, a dip occurs on the high-frequency wings of the LA mode, which indicates that mode coupling exists between the LA mode and the central peak. Similar quasi-elastic scattering was also observed from Raman studies on PbHfO$_3$ [10,26]. Quantitative analysis for this mode coupling phenomena will be presented elsewhere, but it is clear that the central peak intensity increases and the line width, which is usually inversely proportional to the relaxation time of the polar clusters, decreases upon cooling in the cubic phase. This suggests the slowing down behavior of the precursor polar clusters and the increase in their sizes upon cooling.

One method to estimate the relaxation time of the precursor polar clusters is to use the single relaxation time approximation, based on which the LA mode anomalies can be analyzed by using the following equation [21,25].

$$\tau_{LA} = \frac{\Gamma_B - \Gamma_\infty}{2\pi \left(v_\infty^2 - v_B^2\right)}. \qquad (1)$$

In this equation, $v_\infty$ is an unrelaxed, high-frequency limiting Brillouin shift which can be obtained from the linear region at high temperatures, and $\Gamma_\infty$ is the background damping and is estimated to be the average FWHM of the flat region at high temperatures. These values are denoted as flat, solid lines in Fig. 2(a). The calculated inverse of the relaxation time $1/\tau_{LA}$ is plotted in Fig.5. The linear behavior indicates that $\tau_{LA}$ exhibits a diverging behavior with a finite value at $T_H$. This temperature dependence can be fitted by using the following equation.

$$\frac{1}{\pi \tau_{LA}} = \frac{1}{\pi \tau_0} \frac{T - T_0}{T_0} \qquad (2)$$

Here, $T_0$ and $\tau_0$ are fitting parameters. The best-fitted results, shown as a red solid line in Fig. 5, are $T_0$=161 °C and $\tau_0$=0.22 ps. $T_0$ is close to the second antiferroelectric transformation temperature in PbHfO$_3$. This analysis clearly shows that the precursor polar clusters grow significantly, in particular, below approximately 300 °C, reflecting large polarization fluctuations and associated anharmonicity near the phase transition temperature. This kind of precursor dynamics has also been predicted theoretically based on the polarizability model [23] and observed from ferroelectric BaTiO$_3$ and antiferroelectric PbZrO$_3$. It may also be related to the fact that transition entropy associated with transformation at $T_H$ is much larger than that expected for a displacive type transition, thus indicating that at transition point $T_H$ of PbHfO$_3$ we have to deal with a co-existence of the order-disorder and displacive type phase transition mechanism [27].

## IV. CONCLUSION

Elastic properties of high-quality PbHfO$_3$ single crystals were investigated in a wide temperature range by using Brillouin scattering spectroscopy. Two structural phase transitions from low-temperature antiferroelectric to intermediate antiferroelectric and then to paraelectric phases were identified based on abnormal changes in the Brillouin frequency shift and the acoustic damping measured upon heating. Significant softening of the LA mode frequency and the increase in the mode line width in the paraelectric phase strongly suggested the formation and growth of the precursor polar clusters. The estimated relaxation time of the polar clusters exhibited diverging (but limited) behavior indicating slowing down of the precursor dynamics, which was also supported by the observation of the quasielastic central peak. These studies clearly show that the precursor dynamics in perovskite oxides are much more universal than it was suspected. We mean *universality* when microscopic nature of the ferroelectric and/or antiferroelectric phase transitions is discussed.

## ACKNOWLEDGMENTS

This work was supported by the Basic Science Research Program through the National Research Foundation of Korea (NRF) funded by the Ministry of Education, Science and Technology(2013R1A1A2006582).

Figure Captions.

Fig. 1. (Color online) Brillouin spectra of PbHfO$_3$ at three selected temperatures.

Fig. 2. (Color online) Temperature dependence of (a) the LA mode frequency and (b) the full width at half maximum. The inset shows the same data of the TA mode.

Fig. 3. (Color online) Temperature dependence of (a) the LA mode frequency and (b) the full width at half maximum of both PbHfO$_3$ and PbZrO$_3$.

Fig. 4. (Color online) Brillouin spectra of PbHfO$_3$ at three selected temperatures in the cubic phase.

Fig. 5. . (Color online) The inverse of the relaxation time calculated by using the Eq.(1). The red solid line is the best-fitted result by using the Eq.(2).

Figure 1

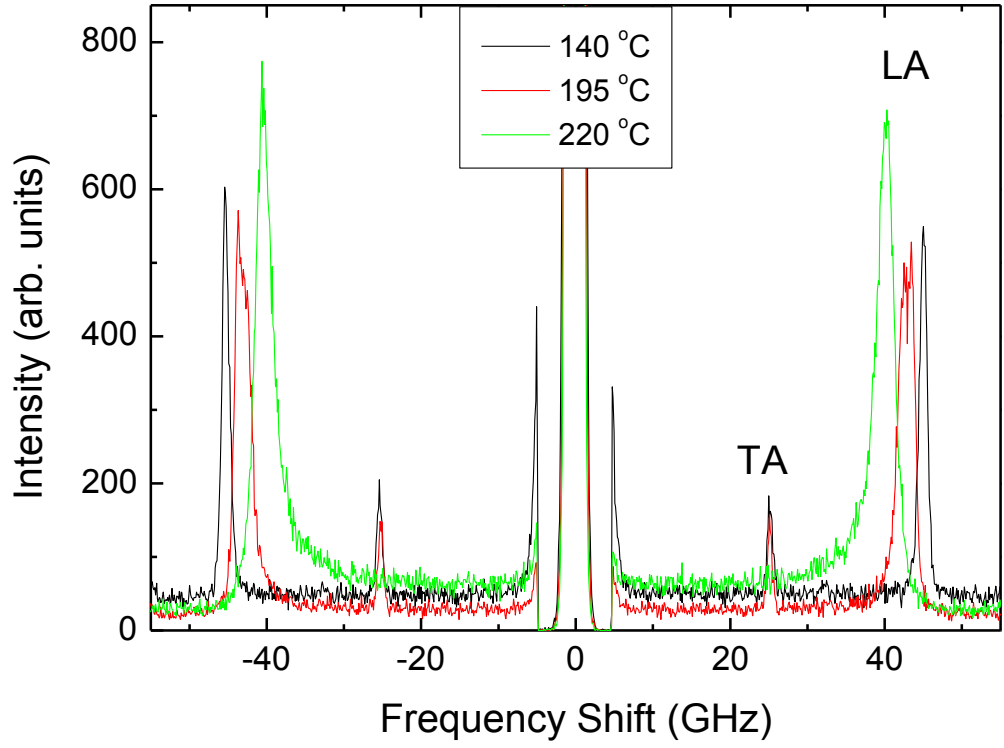

Figure 2

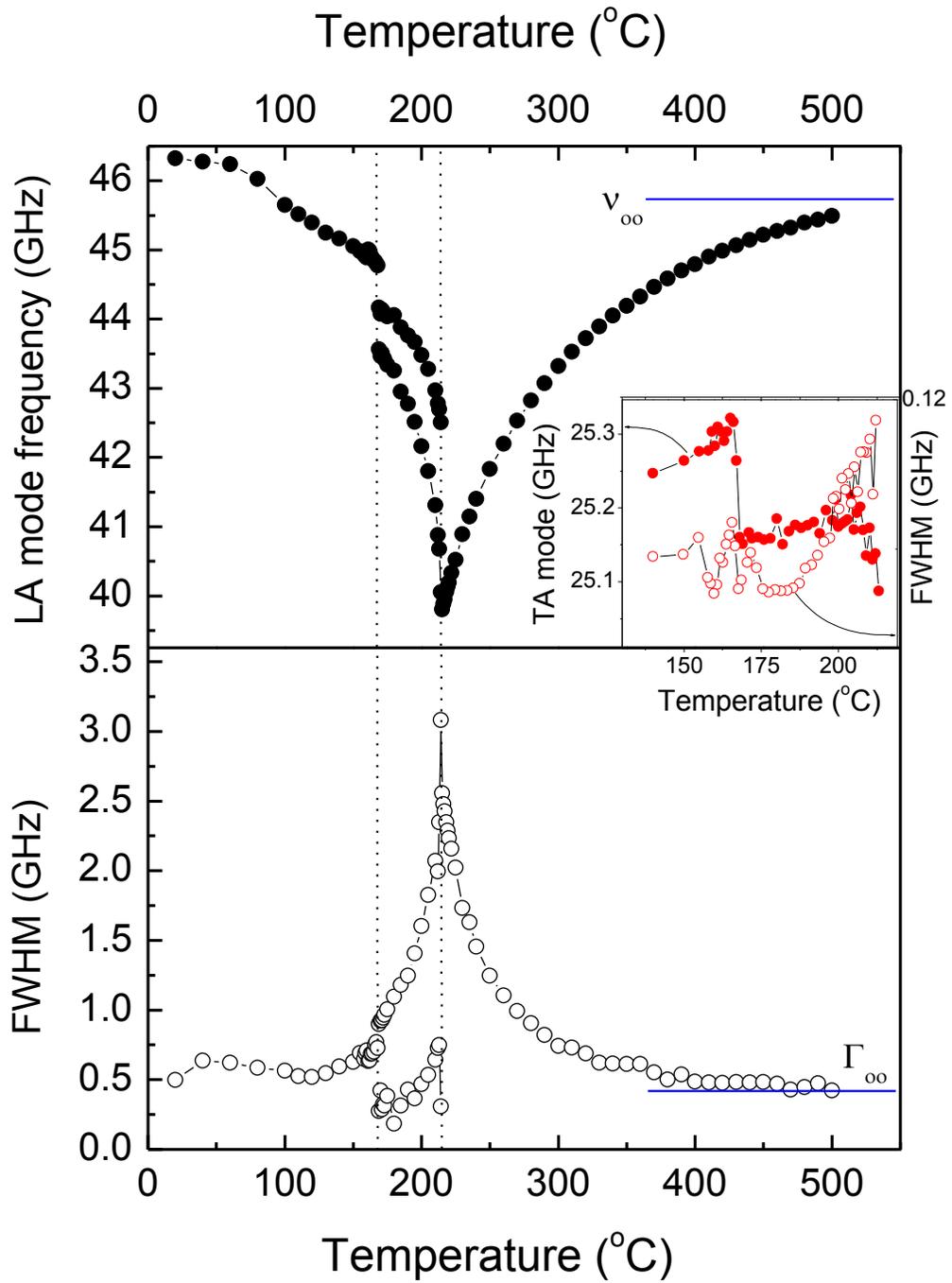

Figure 3

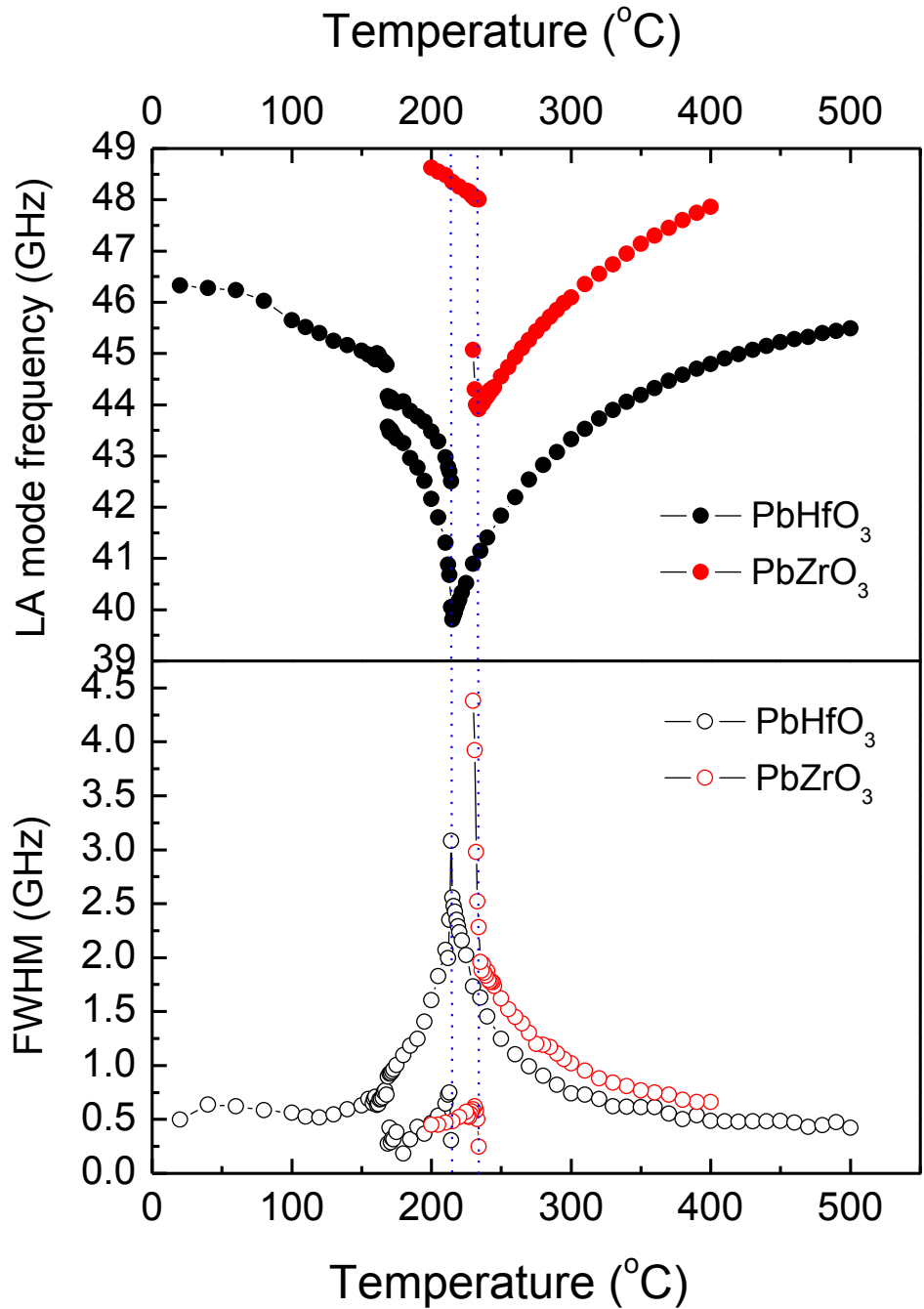

Figure 4.

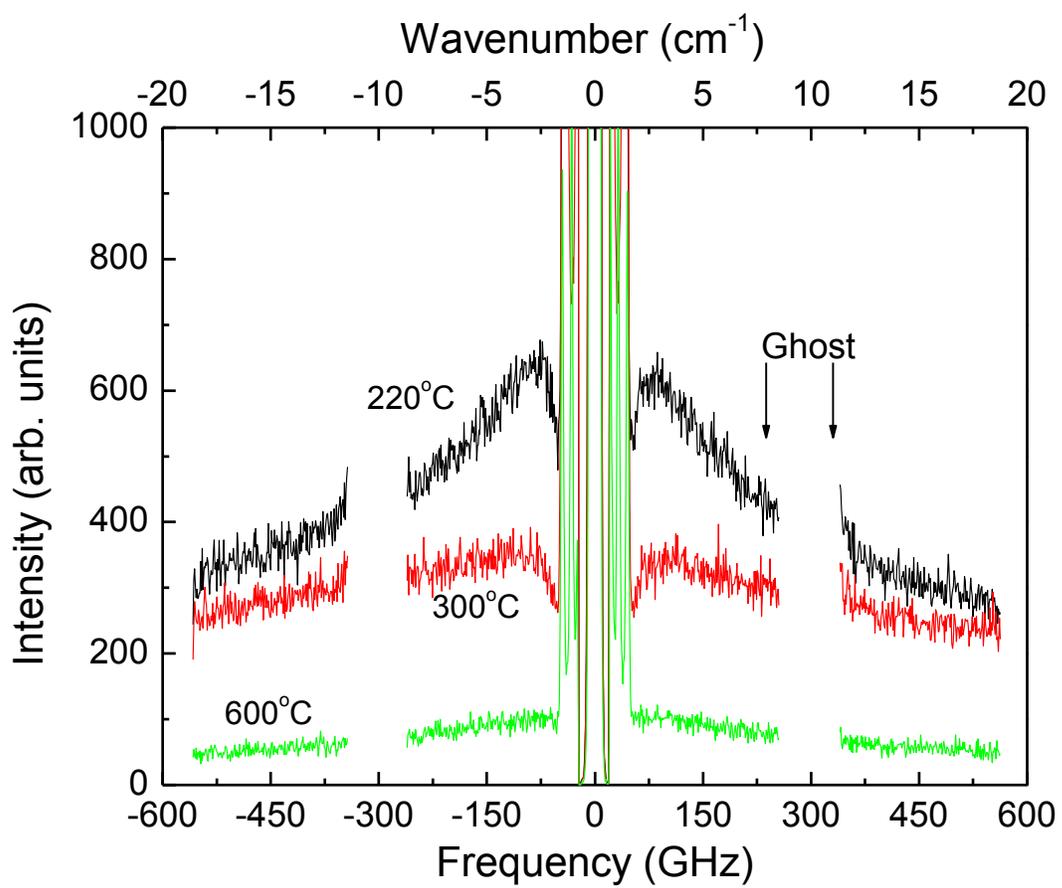

Figure 5

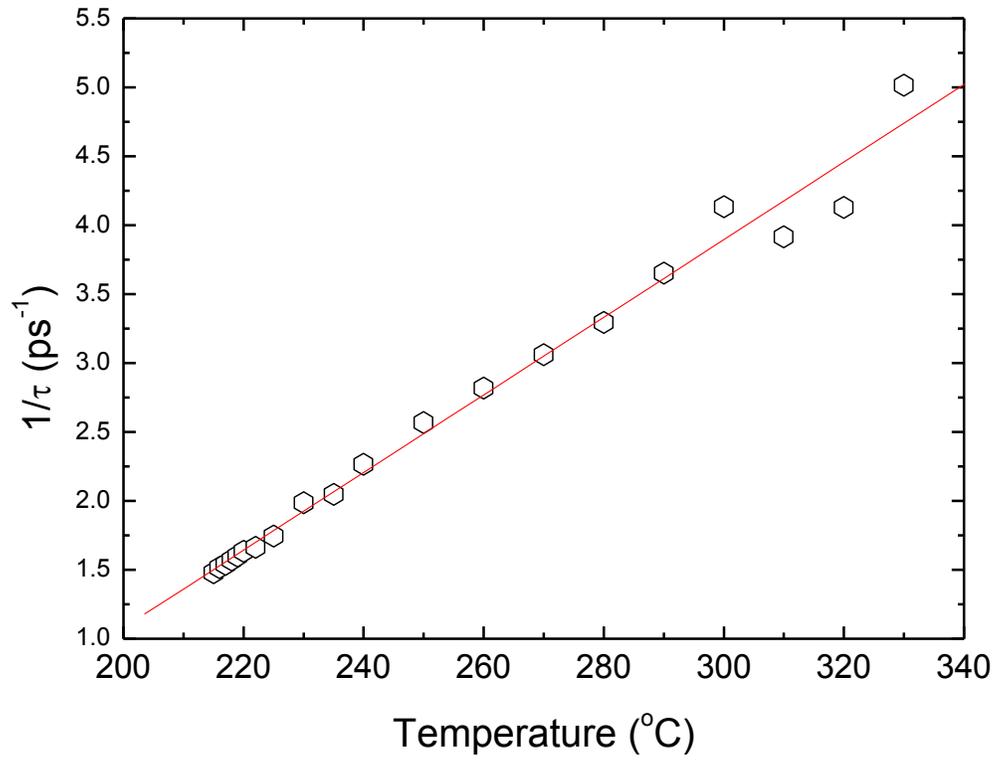